\documentclass[preprint,12pt]{elsarticle}

\usepackage{moreverb,url}
\usepackage{hyperref}
\usepackage{lineno}

\usepackage{microtype}
\usepackage{subfigure}
\usepackage{booktabs} % for professional tables
\usepackage{amsmath}
\usepackage{mathtools}
\usepackage{lipsum}  
\usepackage{adjustbox}
\usepackage{tabularx}
\usepackage{makecell}
\usepackage{siunitx}

\DeclarePairedDelimiter\abs{\lvert}{\rvert}%
\DeclarePairedDelimiter\norm{\lVert}{\rVert}%

\makeatletter
\let\oldabs\abs
\def\abs{\@ifstar{\oldabs}{\oldabs*}}
\let\oldnorm\norm
\def\norm{\@ifstar{\oldnorm}{\oldnorm*}}

\def\printFirstPageNotes{%
  \iflongmktitle
   \let\columnwidth=\textwidth\fi
  \ifx\@tnotes\@empty\else\@tnotes\fi
  \ifx\@cornotes\@empty\else\@cornotes\fi
  \ifx\@elseads\@empty\relax\else
   \let\thefootnote\relax
   \footnotetext{\ifnum\theead=1\relax
      \textit{Email:} \textbf{mateus.diasribeiro@dlr.de, mateusdiasbr@gmail.com}\space\else
      \textit{Email:\space}\fi
     }\fi
  \ifx\@elsuads\@empty\relax\else
   \let\thefootnote\relax
   \footnotetext{\textit{URL:\space}%
     \@elsuads}\fi
  \ifx\@fnotes\@empty\else\@fnotes\fi
  \iflongmktitle\if@twocolumn
   \let\columnwidth=\Columnwidth\fi\fi
  
}

\makeatother

\journal{ }

\let\today\relax
\makeatletter
\def\ps@pprintTitle{%
    \let\@oddhead\@empty
    \let\@evenhead\@empty
    \def\@oddfoot{\footnotesize\itshape
         { } \hfill\today}%
    \let\@evenfoot\@oddfoot
    }
\makeatother

\begin{document}

\begin{frontmatter}


\title{DeepCFD: Efficient Steady-State Laminar Flow Approximation with Deep Convolutional Neural Networks}



\author{Mateus Dias Ribeiro \corref{mycorrespondingauthor} \fnref{ad1,ad2}}
\cortext[mycorrespondingauthor]{Corresponding author:}
%\ead{mateus.dias\_ribeiro@dfki.de}
%\ead{mateusdiasbr@gmail.com}
\ead{mateus.diasribeiro@dlr.de, mateusdiasbr@gmail.com}
%\ead[url]{mateusdiasbr@gmail.com}
\author{Abdul Rehman \corref{} \fnref{ad3}}
\author{Sheraz Ahmed \corref{} \fnref{ad2}}
\author{\\Andreas Dengel \corref{} \fnref{ad2}}
\address[ad1]{German Aerospace Center (DLR), Braunschweig, Germany (current)}
\address[ad2]{German Research Center for AI (DFKI), Kaiserslautern, Germany (previous)}
\address[ad3]{NUST School of Electrical Engineering and Computer Science, Pakistan}

\begin{abstract}
 Computational Fluid Dynamics (CFD) simulation by the numerical solution of the Navier-Stokes equations is an essential tool in a wide range of applications from engineering design to climate modeling. However, the computational cost and memory demand required by CFD codes may become very high for flows of practical interest, such as in aerodynamic shape optimization. This expense is associated with the complexity of the fluid flow governing equations, which include non-linear partial derivative terms that are of difficult solution, leading to long computational times and limiting the number of hypotheses that can be tested during the process of iterative design. Therefore, we propose DeepCFD: a convolutional neural network (CNN) based model that efficiently approximates solutions for the problem of non-uniform steady laminar flows. The proposed model is able to learn complete solutions of the Navier-Stokes equations, for both velocity and pressure fields, directly from ground-truth data generated using a state-of-the-art CFD code. Using DeepCFD, we found a speedup of up to 3 orders of magnitude compared to the standard CFD approach at a cost of low error rates.
\end{abstract}

\begin{keyword}
CFD \sep Deep Learning \sep U-Net


\end{keyword}

\end{frontmatter}

%\linenumbers

\section{Introduction}
\label{sec1}

Computational Fluid Dynamics (CFD) simulations provide detailed description of flow properties of interest for engineering by numerically solving a set of governing Navier-Stokes equations. Their solution, however, can be considerably expensive due to the complexity of the fundamental physics associated with this problem. This expense is a major limitation for the development of products in a wide range of applications, such as aerodynamic design optimization and fluid structure interaction \cite{SWANSON2016102,BATHE2009604}.

\begin{figure}[ht]
\begin{center}
\centerline{\includegraphics[width=0.65\columnwidth]{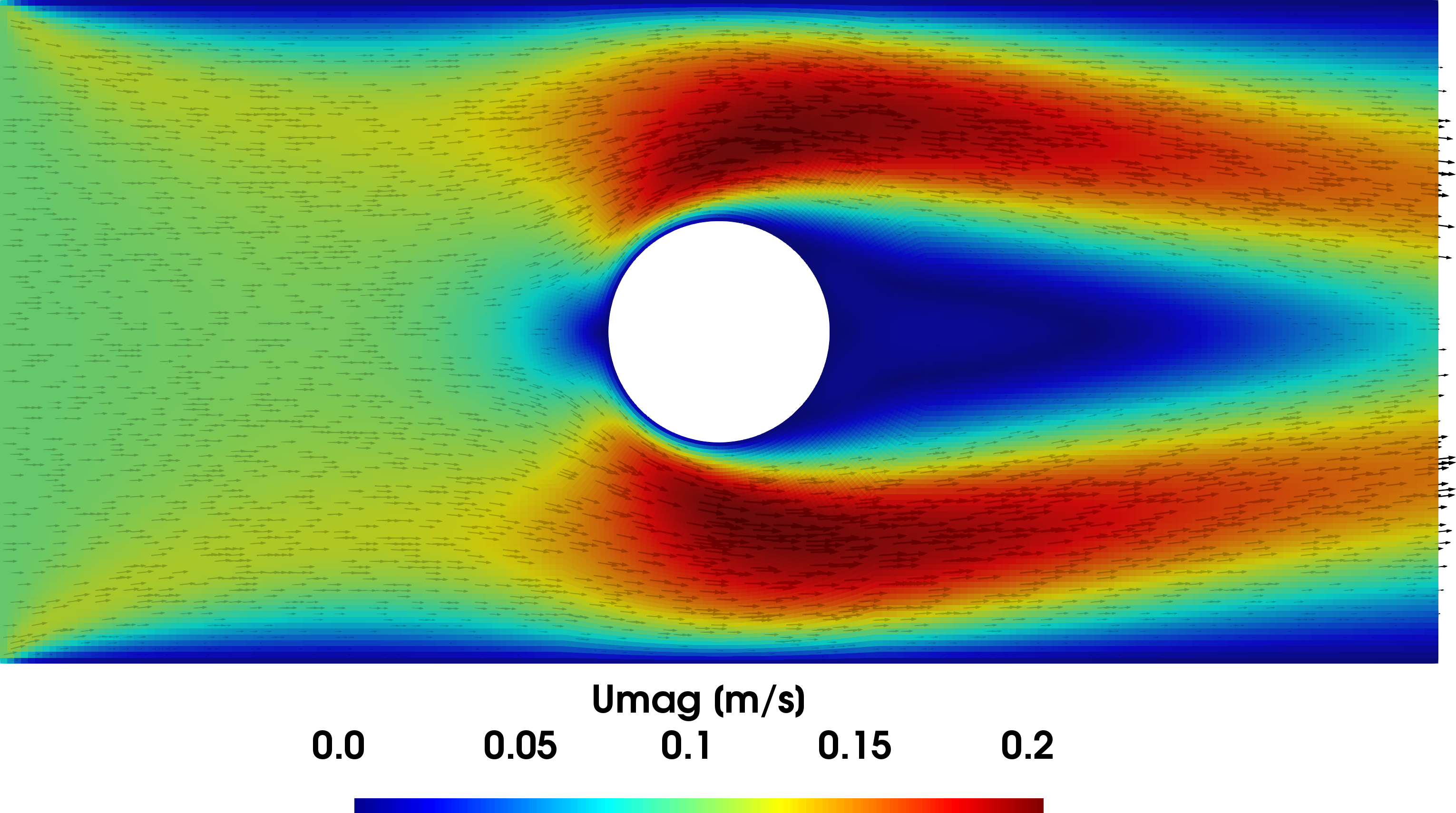}}
\caption{Example of 2D non-uniform steady laminar channel flow around a cylinder.}
\label{fig:fig1}
\end{center}
\vskip -0.4in
\end{figure}

For some engineering applications, the high cost of CFD solutions can be mitigated if certain conditions hold true. For example, if the Reynolds number (relationship between inertial and viscous forces) is low enough, the flow will be laminar, which means that the fluid particles flow along parallel layers with no cross-currents perpendicular to the direction of the flow \cite{pope_2000}, as shown in Figure~\ref{fig:fig1}. Moreover, if the flow achieves a state in which any given property, such as velocity and pressure fields, changes along space but not with time, the problem can be treated as a non-uniform steady laminar flow. In this case, solutions will depend solely on boundary conditions and geometry of the problem. Practical examples in which these conditions can be met in engineering applications are vast, such as flows within nozzles, turbines, compressors, and heat exchangers, as well as in aerodynamic flows under certain conditions, among others \cite{FOLI20061090, FERNANDES2007825, TALUKDAR20083091}. 

During early stages of a product design, such as any of the applications mentioned above, it is important for engineers to test as many hypotheses as possible in order to find a solution that maximizes the performance and/or efficiency of the product being developed. Although state-of-the-art CFD solvers can deliver highly accurate results, the time required to obtain a solution inhibits extensive iterative design due to the expensive resources necessary to run these codes. Therefore, the use of data-driven machine learning approaches to generate an accurate approximation of these simulations using a fraction of their resources is very appealing \cite{portwood2019turbulence}. The potential of these approximated solutions for accelerating the results at a cost of low error rates \cite{kddpaper} can lead to more efficient development of products dependent on CFD research. Finally, these approaches may also be implemented together with physics-driven techniques in order to bound the model prediction within the physical constraints of the problem \cite{chen2018neural, RAISSI2019686, portwood2019turbulence}. 

The literature on data-driven methods for CFD provides several contributions to the field of machine learning models for fluid flow predictions. Pioneering works from \cite{SARGHINI200397} and \cite{kddpaper} show how neural networks can be employed for deriving closure terms for turbulence modeling and for predicting the velocity magnitude field in steady-state flows. Moreover, other recent works from distinct sources provide further contributions to relevant fields in a wide range of applications, such as physics-informed neural networks, airfoil design optimization, acceleration of sparse linear system solutions, etc  \cite{ling_kurzawski_templeton_2016,Lui_2019,tompson2016accelerating,dias2019data}. 

In this paper, we present DeepCFD: a deep learning model based on CNN architectures to provide an approximated solution for both velocity and pressure fields by feeding it ground-truth data of a channel flow around randomly shaped obstacles, generated by a state-of-the-art CFD solver. The following summarizes the main contributions of this work: 

\begin{enumerate}
   \item {We propose a CNN-based surrogate CFD model for 2D non-uniform steady-state laminar flows that provides a solution with up to 3 orders of magnitude speedup at a cost of low error rates.}
   
    \item {We extend previous efforts from other researchers, such as in Guo et al.~\citep{kddpaper}, by providing complete solutions of all components of the velocity field and the pressure field. A complete solution for velocity and pressure is essential for engineers to develop products that interact with a given flow field, or to account for the transport of a scalar of interest in industrial flows. }
    
    \item {Provide the code and dataset used in this work for the general public to contribute with further expansion of the field of data-driven models for CFD: \url{https://github.com/mdribeiro/DeepCFD} }

\end{enumerate}

\section{Methodology}
\label{sec2}

In this section, the traditional CFD approach is presented, followed by the proposed surrogate DeepCFD approach based on CNN architectures. The traditional CFD approach provides the ground-truth data for training DeepCFD.

\subsection{CFD Approach: OpenFOAM}
\label{sec21}

The incompressible transient two dimensional Navier-Stokes equations for mass (\ref{eq:eq1}) and momentum (\ref{eq:eq2}) conservation read:

\begin{equation}
\label{eq:eq1}
\nabla \cdot \mathbf u = 0
 \end{equation}

\begin{equation}
\label{eq:eq2}
 \rho\left(\frac{\partial}{\partial t}+\mathbf u\cdot{\rm div}\right)\mathbf u=-\nabla p+\nabla\cdot\mathbf \tau + \mathbf f
 \end{equation}

\noindent in which u is the velocity field (with x and y components for 2 dimensional flows), $\rho$ is the density, $p$ is the pressure field, $\tau$ is the stress tensor, and $f$ represents body forces, such as gravity. 
 
If a non-uniform steady-state flow condition is assumed, the accumulation term (time $t$ dependence term) is dropped, and the momentum equation can be rewritten for velocity components $u_x$ (\ref{eq:eq3}) and $u_y$ (\ref{eq:eq4}) as:

\begin{equation}
\label{eq:eq3}
u_{x}{\frac {\partial u_{x}}{\partial x}}+u_{y}{\frac {\partial u_{x}}{\partial y}}=-\frac{1}{\rho}{\frac {\partial p}{\partial x}} + \nu \nabla^2 u_{x} + g_{x}
\end{equation}

\begin{equation}
\label{eq:eq4}
u_{x}{\frac {\partial u_{y}}{\partial x}}+u_{y}{\frac {\partial u_{y}}{\partial y}}=-\frac{1}{\rho}{\frac {\partial p}{\partial y}} + \nu \nabla^2 u_{y} + g_{y}
\end{equation}
 
\noindent in which $g$ represents the gravitational acceleration and $\nu$ the dynamic viscosity of the fluid. The terms on the left-hand side of these equations account for the convective transport, whereas the terms on the right-hand side account for the pressure coupling and diffusive transport.

The above equations are solved numerically using the \textit{simpleFoam} solver from \textit{OpenFOAM} \cite{ofpaper}, an extensively validated C++ written framework for the numerical solution of partial differential equation systems. The solver is based on the "Semi-Implicit Method" or \textit{SIMPLE} algorithm \cite{patankar1980numerical}, which obtains the solution of velocity and pressure fields by iteratively updating initial guesses by correction terms based on the mass and momentum conservation equations. The discretized momentum equation is solved implicitly in two separate steps, with the first providing an explicit solution for the velocity field based on the current pressure field, and the second providing an implicit correction for the pressure field using information from the previous step. This procedure is repeated until convergence is achieved, which can take a considerable amount of time if the required level of accuracy is high. Moreover, if high spatial resolution is desired, more grid elements will be necessary to perform the simulation. Since the discretized equations need to be solved at each grid element, the number of evaluations at each iteration can become very large.

\subsection{Machine Learning Approach: DeepCFD}
\label{sec22}

In order to overcome the issues of the CFD approach, mainly regarding computational cost and solution time, DeepCFD is proposed to leverage from deep convolutional neural networks (DCNN) in order to create a surrogate model to provide efficient velocity and pressure solutions for steady-state laminar flows. 

\subsubsection{Convolutional Neural Networks}
\label{sec221}

Convolutional Neural Networks (CNNs) have proven great capability of learning important features from images at the pixel level in order to make useful predictions for both classification and regression problems \cite{deepdoc, airfoilcnn}. Another advantage of this approach, compared to conventional fully-connected layer networks, lies in the fact that convolutions provide weight-sharing and sparse connectivity \cite{Goodfellow-et-al-2016}. These properties enable more efficient memory usage to learn the necessary information needed to create a surrogate model to reconstruct an approximation of whole velocity and pressure fields from a given set of boundary conditions. 

In the case of steady-state flows, the solution will be dependent solely on the boundary conditions, such as the geometry of the problem. Therefore, the input layer of the CNN model needs to provide information regarding the geometric information of the flow. For this task, we use the signed distance function (SDF) proposed by \cite{kddpaper}, which provides the distance of individual points in the CFD grid to a given surface (such as the cylinder in Figure~\ref{fig:fig1}), given by: 

\begin{equation}
\label{eq:eq5}
\displaystyle SDF(x)={\begin{cases}d(x,\partial \Omega ){\mbox{ if }}x\in \Omega \\-d(x,\partial \Omega ){\mbox{ if }}x\in \Omega ^{c}\end{cases}}
\end{equation}

\noindent where $\Omega$ is a subset of a metric space, X, with metric, d, and $\partial \Omega$ is the boundary of $\Omega$. For any x $\in$ X:

\begin{equation}
\label{eq:eq6}
\displaystyle d(x,\partial \Omega ):=\inf _{y\in \partial \Omega }d(x,y)
\end{equation}

\noindent where $inf$ denotes the infimum. Grid positions inside the obstacle's interior ($\Omega ^{c}$) are assigned negative distances.

\begin{figure}[ht!]
\begin{center}
\centerline{\includegraphics[width=0.8\columnwidth]{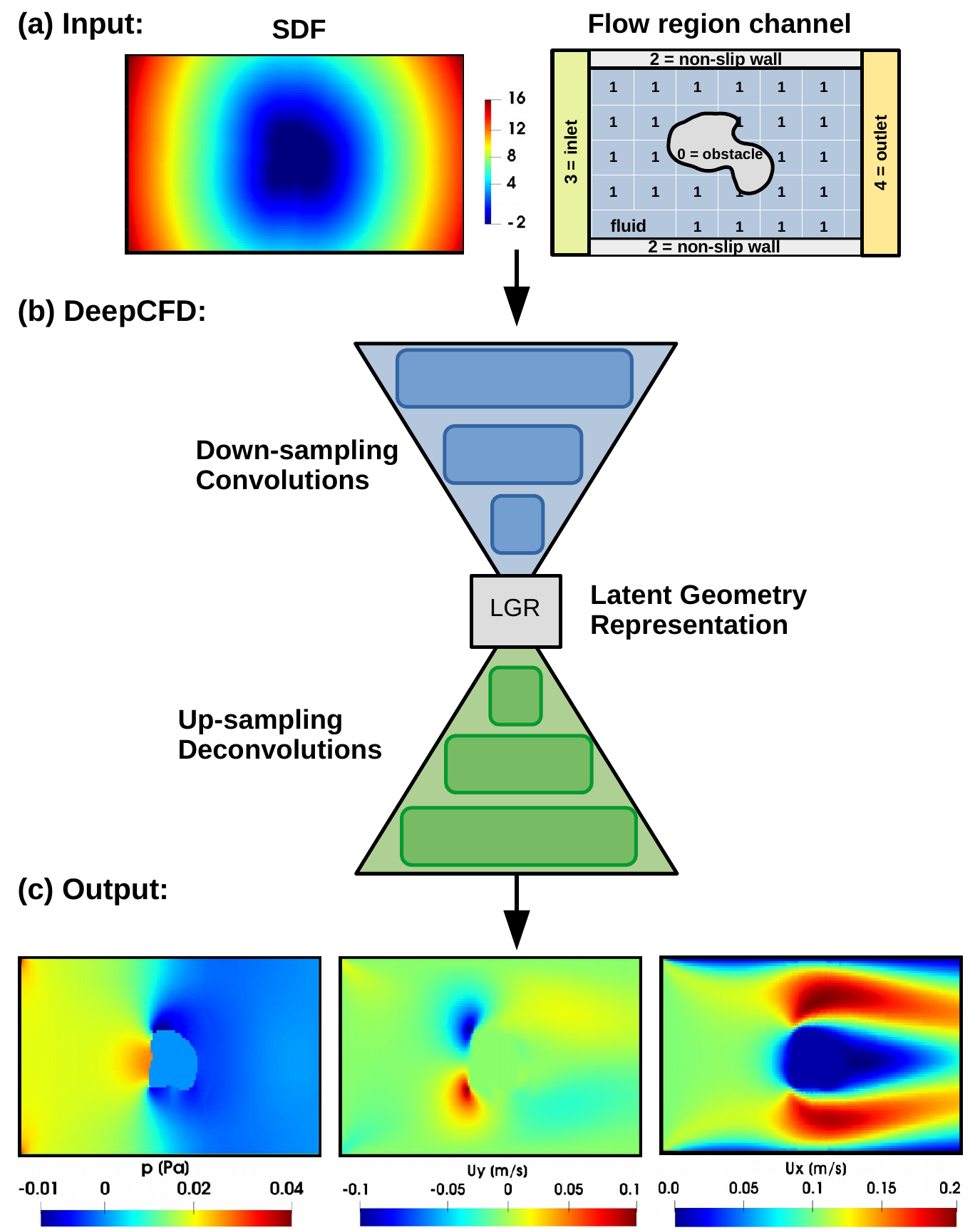}}
\caption{CNN learning approach. (a) Input channels with SDF and multi-class labeling of flow regions. (b) Down-sampling convolutional operations create a latent representation of the flow geometry from the input. (c) Up-sampling deconvolutions map the LGR to variables of interest.}
\label{fig:fig2}
\end{center}
\vskip -0.2in
\end{figure}

Moreover, a multi-class channel with information about the flow region in 5 different categories (0 for the obstacle, 1 for the free flow region, 2 for the upper/bottom no-slip wall condition, 3 for the constant velocity inlet condition, and 4 for the zero-gradient velocity outlet condition) is provided. The employment of consecutive down-sampling convolutional operations encode the given input into a latent geometry representation (LGR), as illustrated in Figure~\ref{fig:fig2}-b. Input channels are illustrated in Figure~\ref{fig:fig2}-a.

After encoding the geometric information, one can use transposed convolutions (also known as "deconvolutions") to find a mapping between the LGR and any variable of interest, such as the velocity field (Ux and Uy) and the pressure field (p). This can be done by performing upsampling deconvolution operations from the LGR encoding until the original CFD dimension size is achieved, with the number of output channels equal to the number of variables of interest (Figure~\ref{fig:fig2}-c). The goal of the training is to minimize the total error function based on a given criterion (e.g. absolute mean error or mean squared error) between the CNN model and the ground-truth CFD results. 

\subsubsection{Neural Network Architectures}
\label{sec222}

\begin{figure}[ht!]
\begin{center}
\centerline{\includegraphics[width=\columnwidth]{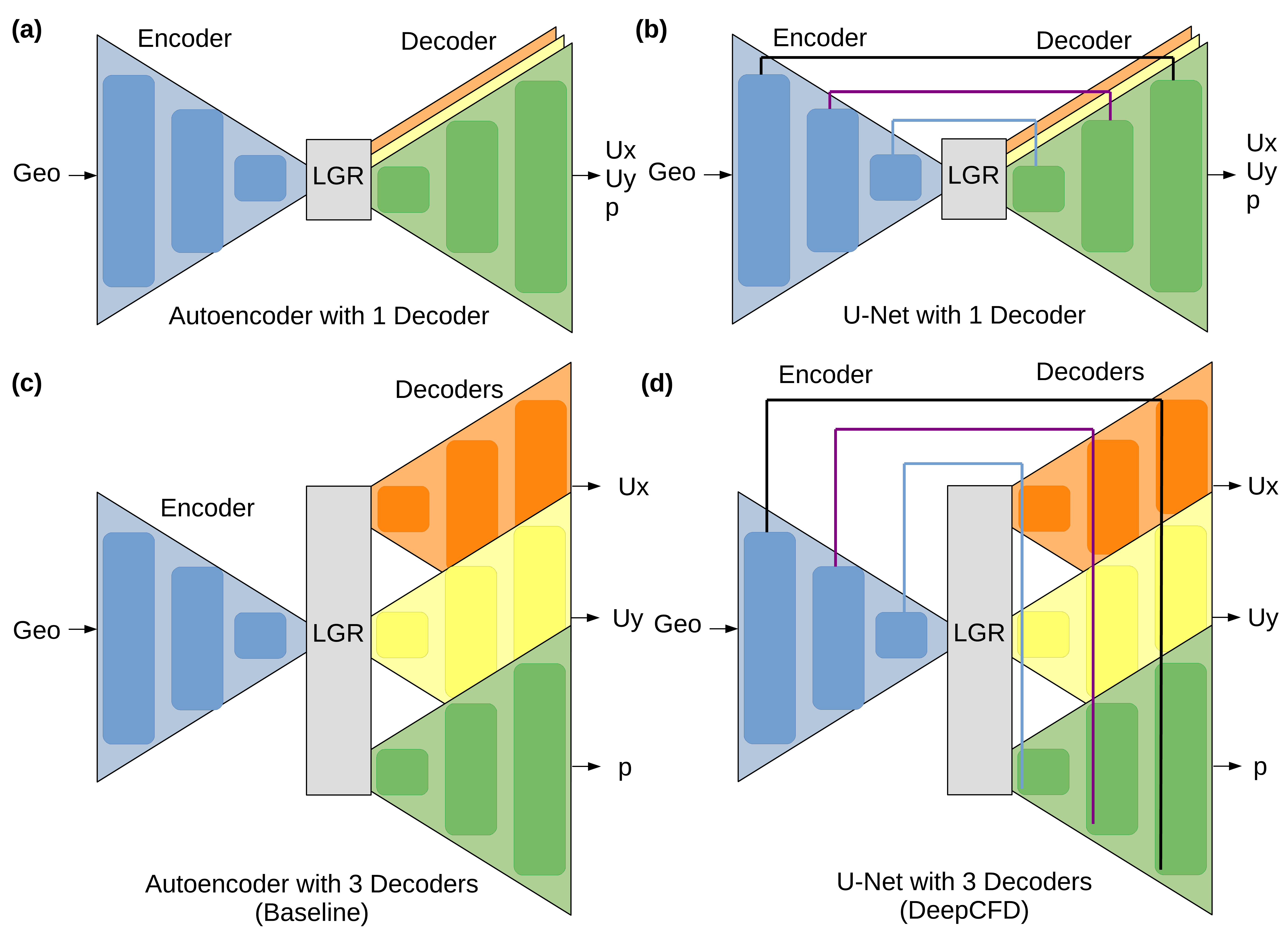}}
\caption{Convolutional neural network architectures and respective variants investigated in this work.}
\label{fig:fig3}
\end{center}
\vskip -0.2in
\end{figure}

In this work, a modified version of the architecture used by \cite{kddpaper} is proposed. In their approach, a down-sampling encoder network compresses the geometric information (SDF) into a reduced dimension latent space or latent geometry representation (LGR), followed by an up-sampling decoder network that maps the LGR encodings back to data space (velocity components). For its similarity to autoencoder networks, this variant is named here as "Autoencoder" or simply "AE", and is considered our baseline. Moreover, each output variable is mapped from the LGR encoding using separate decoder networks. In our proposed approach, since the model must also be able to provide a solution for the pressure field, an U-Net architecture is employed. This kind of network was proposed by \cite{ronneberger2015u} for the task of segmentation of medical images, and we show that such networks can also be successfully applied here to find an efficient mapping between geometry and steady-state flow solution of coupled pressure-velocity fields. As shown in the schematic representation provided in Figure~\ref{fig:fig3}, both baseline and DeepCFD models use the architecture variants with multiple decoders: AE-3 (c) and UNet-3 (d). For completeness, the effect of disentangling the model outputs in separate decoders was put to test by including architectures with 1 decoder for both Autoencoder and U-Net variants: AE-1 (a) and UNet-1 (b).

In order to find the ideal setup for the architectures mentioned above, a set of hyper-parameters were considered for investigation. These parameters are summarized in Table~\ref{tab:table1}, and include three different learning rates, three different CNN filter sizes (kernel), and a number of encoder/decoder blocks (3, 4, or 5) with varying number of filters (ranging from 8 to 64). Each encoder/decoder block contains two convolutional layers and the number of filters used in the encoder network is reversed for the decoder network. For example, if a $\left[16,32,64\right]$ filter configuration is used in the encoder, a $\left[64,32,16\right]$ configuration is used in the decoder. Max pooling operations are performed for each layer and the ReLU activation function is applied to the layer outputs. Furthermore, batch and weight normalization may or may not be performed. Therefore, each architecture can assume 3x3x3x2x2 = 108 different configurations, which were extensively evaluated during hyper-parameter search.

\begin{table}[ht!]
\caption{Set of parameters considered for hyper-parameter search.}
\label{tab:table1}
\vskip 0.15in
\begin{center}
\begin{small}
\begin{sc}
\begin{tabularx}{.75\columnwidth}{lccc}
\toprule
Parameters \\
\midrule
Learning rate   & 1e-3 & 1e-4 & 1e-5 \\
Kernel    & 3 & 5 & 7 \\
Filters    & $16,32,64$ & $8,16,32,32$ & $8,16,16,32,32$ \\
\midrule
Norm & & & $ $  \\
\midrule
Batch  & On & Off  \\
%\midrule
Weight    & On & Off  \\
\bottomrule
\end{tabularx}
\end{sc}
\end{small}
\end{center}
\vskip -0.1in
\end{table}

\section{Problem Setup}
\label{sec3}

\begin{figure}[ht!]
\begin{center}
\centerline{\includegraphics[width=0.7\columnwidth]{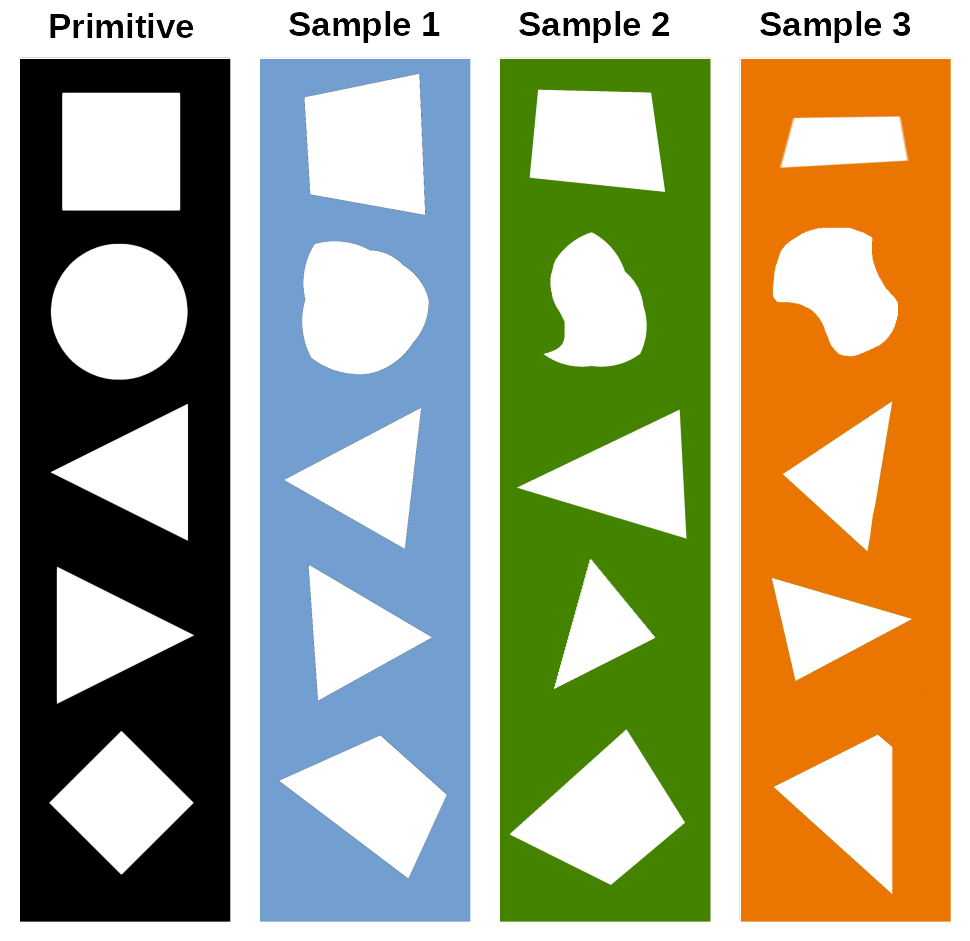}}
\caption{Primitive and derived obstacle forms.  }
\label{fig:fig4}
\end{center}
\vskip -0.2in
\end{figure}

In this section, the setup for generating the ground-truth CFD data and the training procedure of DeepCFD are described. A diverse dataset of 2D steady-state channel flow around random shaped obstacles is simulated with the \textit{simpleFoam} solver in more than 1000 instances. For each instance, a new obstacle mesh is used by sampling it from a random shape generator. The generator creates obstacle shapes based on five different primitive shapes (circle, square, forward-facing triangle, backward-backing triangle, and rhombus) by randomly shifting the points used to construct these shapes in a given range of directions. Figure~\ref{fig:fig4} illustrates some examples of the random samples generated that were used as obstacles for the 2D channel flow dataset.

\begin{figure}[ht!]
\begin{center}
\centerline{\includegraphics[width=1.0\columnwidth]{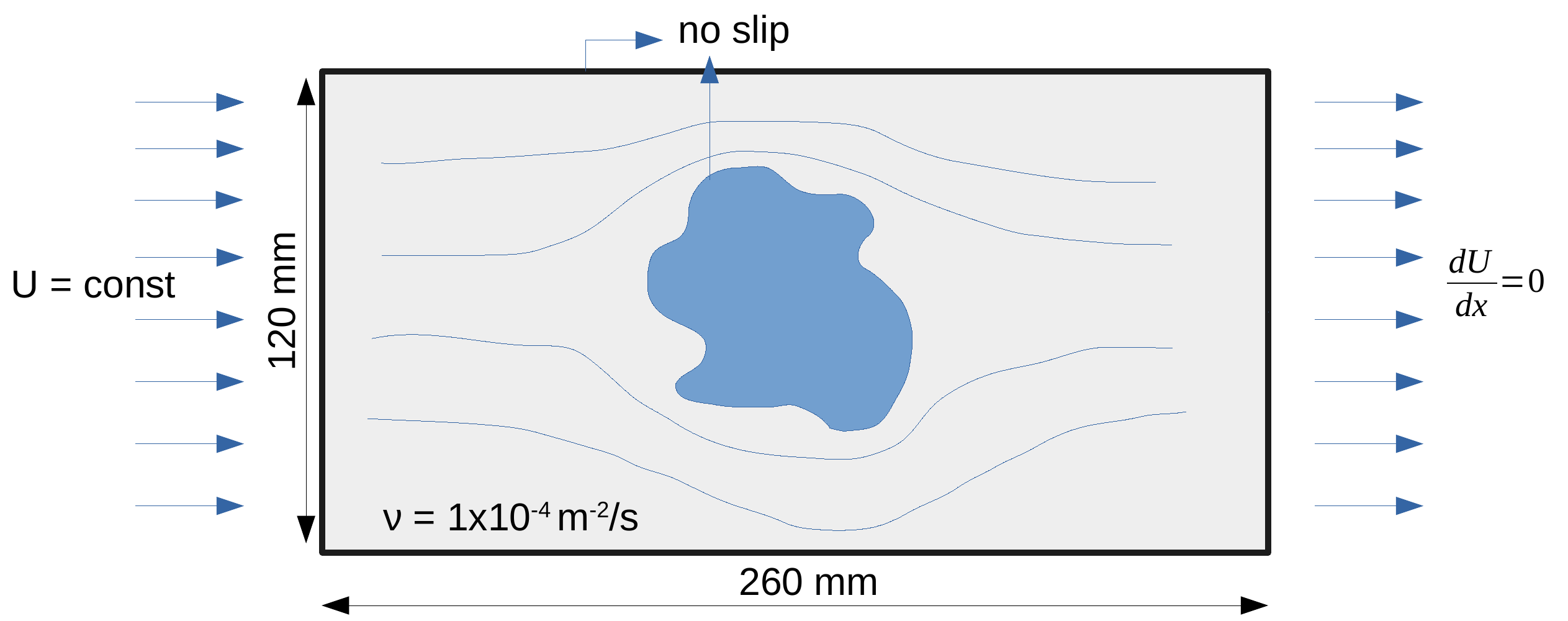}}
\caption{Simulation domain and boundary condition details.  }
\label{fig:fig5}
\end{center}
\vskip -0.4in
\end{figure}

The domain dimensions are 260~mm in the stream-wise direction and 120~mm in the direction perpendicular to the flow. The number of grid elements varies according to the shape used, but with a base cell size of around 1~mm, the average cell count is about 30,000. Boundary conditions are kept fixed, with a constant radial velocity of 0.1~m/s on the inlet (left wall), a zero-gradient condition on the outlet (right wall), and no-slip boundary condition on the top/bottom and obstacle walls, as shown in Figure~\ref{fig:fig5}. Furthermore, the laminar dynamic viscosity is set to $\num{1e-4}$~$m^2/s$, central differencing schemes (CDS) were used for the discretization of both convective and diffusive terms of the momentum equation, and simulations were run on a single core of an Intel Xeon E-2146G processor.

A shell-script was written to automatize the process of dataset generation by calling the random geometry generator, the \textit{OpenFOAM} internal mesher \textit{blockMesh}, and the solver \textit{simpleFoam}. The results for all components of the velocity field and the pressure field were saved for each sample together with the cell location information of the entire computational grid. In the next element of the pipeline, the CFD results of all samples are converted into numpy arrays, so that SDF and multi-class region channels can be calculated and easily inputted to the PyTorch workflow used to train the DeepCFD model.

Regarding the learning process, the AdamW optimizer was employed with a batch size of 64 and weight decay was set to 0.005. The network was trained on an Nvidia Tesla V100 SXM2 GPU using a 70~\%-30~\% split for training and testing. The loss function combines the errors of all three outputs, velocity components Ux and Uy and the pressure. For the velocity components, a mean squared error function was used, whereas for the pressure a mean absolute error was employed. This choice of loss functions was based on extensive experiments, which showed considerably better convergence when L2-norm was applied to the velocity components and L1-norm used for the pressure. A possible justification for that may be related to characteristics of the pressure-velocity coupling in the momentum equation. In certain situations, the coupling can become unstable and under-relaxation of the pressure becomes necessary to improve the robustness of the pressure solution. Therefore, the use of a loss function more resistant to outliers, such as the L1-norm, may provide a better coupling between velocity and pressure for the reconstruction by the machine learning approach. Care was taken to normalize each individual loss in order not to bias the model towards optimizing a particular output in detriment of the others.

\section{Results}
\label{sec4}

In this section, the capability of the proposed model in providing efficient approximations of steady-state laminar flow solutions is demonstrated. First, the model optimization procedure via hyper-parameter search is described, and the test error curves of DeepCFD are plotted against the ones of the baseline model \citep{kddpaper}. Furthermore, qualitative and quantitative analyses of the results are provided together with relevant discussion about the model accuracy and performance in comparison with the baseline \citep{kddpaper} and with the standard CFD approach.

\subsection{Model Optimization}
\label{sec41}

The 108 different parametric configurations introduced in Table~\ref{tab:table1} of section~\ref{sec222} were extensively tested for each of the four architectures (Figure~\ref{fig:fig3}) considered in this work, yielding a total of 432 experiments. Each experiment was conducted 5 times for 1000 Epochs in order to ensure that results are reproducible, and each model was trained with 700~samples and tested with 300~samples, generated using the setup described in section~\ref{sec3}. The average and standard deviation of the mean squared error results on the test set for the best model from each architecture are shown in Table~\ref{tab:table2}:

\begin{table}[ht!]
\caption{Model performance comparison between best baseline and DeepCFD models. Additional 1 decoder configuration for each case was added to test effect of multiple decoders.}
\label{tab:table2}
\vskip 0.15in
\begin{center}
\begin{small}
\begin{sc}
\begin{tabular}{lcc}
\toprule
    &     \text{n = 5 samples} &   \\
\midrule
MSE &    AE-1 &  Baseline  \\
\midrule
Ux    &   \text{2.1513 $\pm$ 0.1688} &  \text{1.7854 $\pm$ 0.1175}  \\
Uy   & \text{0.6270 $\pm$  0.0611} &   \text{0.2956 $\pm$  0.0045}   \\
p      & \text{1.7198 $\pm$ 0.0052} &   \text{1.2125 $\pm$ 0.0150}  \\
Total   & \text{4.4981 $\pm$ 0.1753} &     \text{3.2935 $\pm$ 0.1171}        \\
\midrule
MSE &   UNet-1 & DeepCFD   \\
\midrule
Ux      & \text{1.1169 $\pm$ 0.1393} &  \textbf{ 0.7730 $\pm$ 0.0897  }  \\
Uy &   \text{0.3326 $\pm$  0.0121} &   \textbf{  0.2153 $\pm$  0.0186 }   \\
p      & \text{1.4708 $\pm$ 0.0045} &   \textbf{  1.0420 $\pm$  0.0431 }  \\
Total   & \text{2.9203 $\pm$ 0.1520} &     \textbf{ 2.0303  $\pm$  0.1360  }      \\
\bottomrule
\end{tabular}
\end{sc}
\end{small}
\end{center}
\vskip -0.1in
\end{table}

The DeepCFD model (UNet-3 architecture) outperformed the baseline and all other architectures in regard to the mean squared error (MSE) on the test set for all variables of interest (Ux, Uy, and p). The other exact parameters of this selected configuration include a learning rate of $\num{1e-3}$, a kernel size of 5, a total of eight convolutional layers for each encoder/decoder network (using the $\left[8,16,32,32\right]$ number of filters configuration), and both batch and weight normalization turned off. Our hypothesis that the skip-connections of the U-Net network contributes to more accurate reconstructions than the baseline model by providing direct connections between the encoded geometry features and the decoder layers for each variable was confirmed. Moreover, the proposed DeepCFD architecture with separate decoders also shows significant advantage in comparison to the single decoder model. The total MSE using the DeepCFD network was about 70~$\%$ of the one given by UNet-1. One possible explanation for the observed differences is the modeling complexity of the pressure-velocity coupling in the momentum equation, with a non-linear convection term containing a velocity-velocity coupling and a linear pressure-velocity coupling. Therefore, better approximation of the steady-state flow can be obtained with separate decoder networks because each contains its own set of learnable parameters.

\subsection{Error curves on the Test-set  }
\label{sec42}

\begin{figure}[ht!]
\begin{center}
\centerline{\includegraphics[width=1\columnwidth]{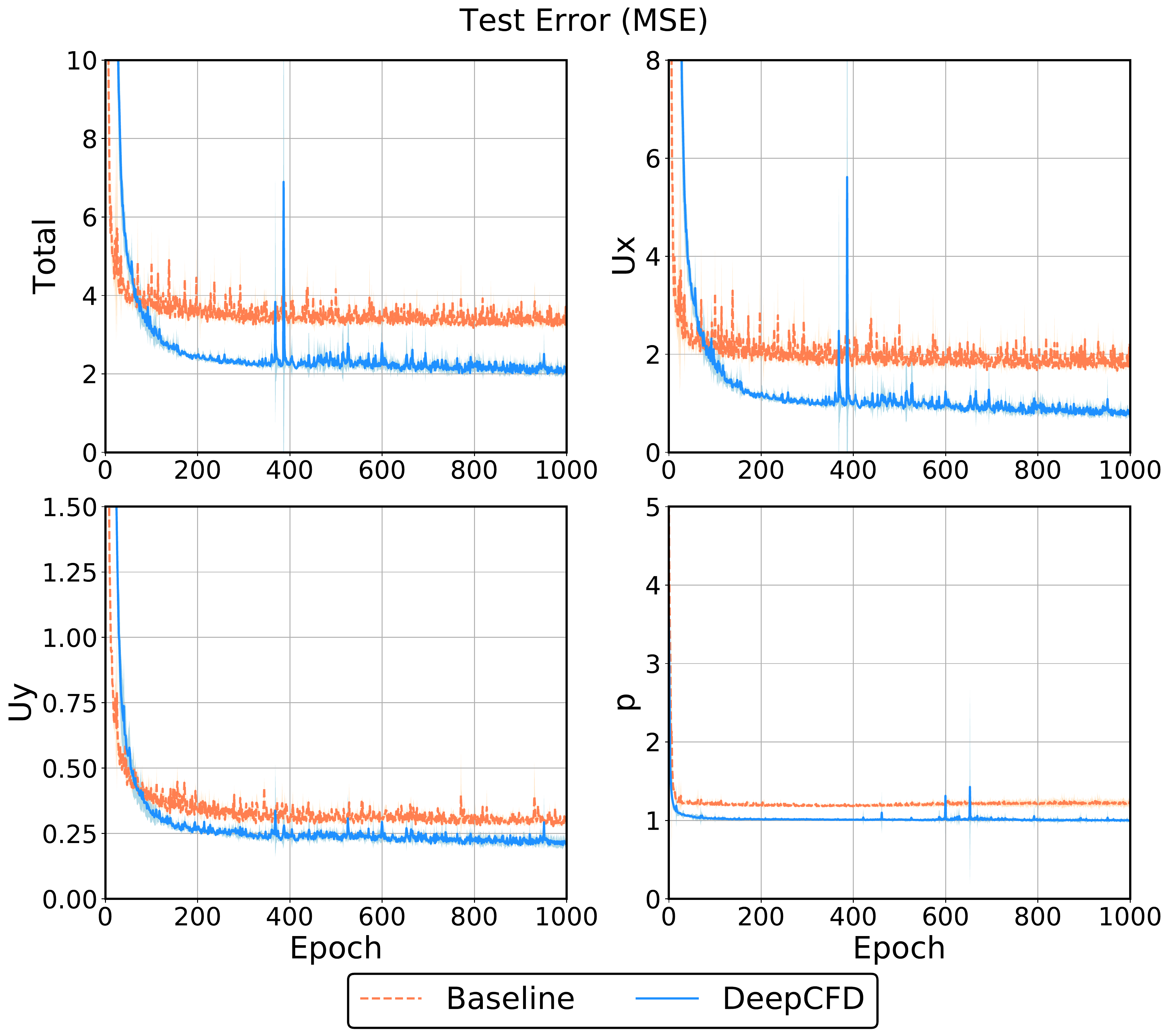}}
\caption{Test MSE vs Epoch curves for the total error, horizontal and vertical velocity components, and pressure.  }
\label{fig:fig6}
\end{center}
\vskip -0.2in
\end{figure}

Test MSE versus Epoch curves for all variables and the total combined error are shown for 1000 Epochs in Figure~\ref{fig:fig6}. Each plot includes the results from the baseline model (dashed orange line) and from the DeepCFD model (continuous blue line). In addition to the central tendency (mean curves), the observed standard deviations are also plotted as shaded regions around the mean curves.

Although the baseline model shows slightly better initial performance in predicting both components of the velocity field (Ux and Uy), it takes about 100 Epochs for DeepCFD to achieve the same performance while it steadily continues to improve. After 100 Epochs, overall test error ratio between baseline and DeepCFD is already 1.2, with this figure increasing to 1.6 times at 1000 Epochs. Regarding the accuracy of the pressure prediction, the proposed model outperforms the baseline from the beginning, achieving an MSE about 17~$\%$ smaller than the baseline total MSE and remaining around 1 during almost the entire training time.

\subsection{Flow Visual Inspection}
\label{sec43}

\begin{figure}[ht!]
\begin{center}
\includegraphics[width=1\columnwidth]{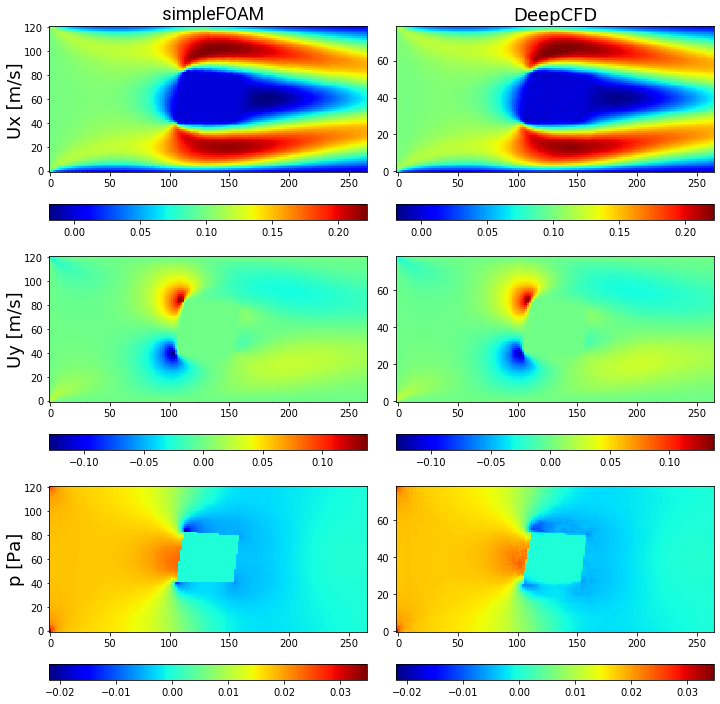}
\caption{Comparison between ground-truth CFD (simpleFOAM) and DeepCFD prediction, showing both velocity components and pressure fields in flow around square based shape.}
\label{fig:fig7}
\end{center}
%\vskip -0.2in
\end{figure}

Qualitative plots of flow reconstructions provided by DeepCFD on test-set samples are presented and compared to the ground-truth solution. For instance, results of both velocity and pressure fields from the flow around an obstacle based on the square primitive form are shown in Figure~\ref{fig:fig7}. The left column refers to the ground-truth CFD data (simpleFOAM), whereas the right column presents the DeepCFD results. It is good practice to show all components of vector fields, such as the velocity, in order to provide information not only about the magnitude but also about the direction of the flow. Therefore, the first two rows show the horizontal (x) and vertical (y) components of the velocity field (U), while the third row provides the pressure field (p) results. 

The square shaped obstacle induces a high pressure region on its frontal edge, as well as forces the flow to separate right after the fluid reaches the frontal vertices of the obstacle. This behavior is also well captured by the DeepCFD model in both velocity and pressure plots.

\begin{figure}[ht!]
\begin{center}
\includegraphics[width=1\columnwidth]{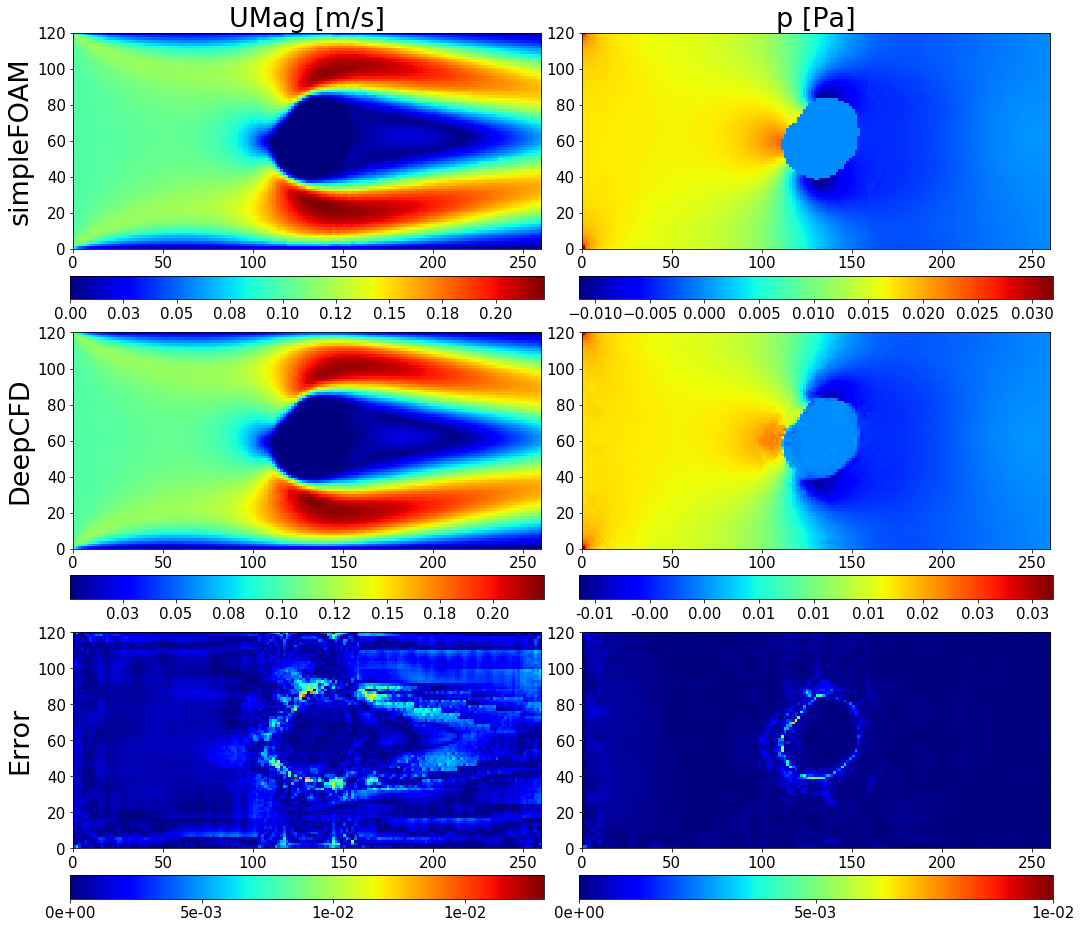}
\caption{Comparison between ground-truth CFD (simpleFOAM) and DeepCFD prediction, showing velocity magnitude, pressure, and absolute error in flow around circle based shape.}
\label{fig:fig8}
\end{center}
%\vskip -0.2in
\end{figure}

\begin{figure}[ht!]
\begin{center}
\includegraphics[width=1\columnwidth]{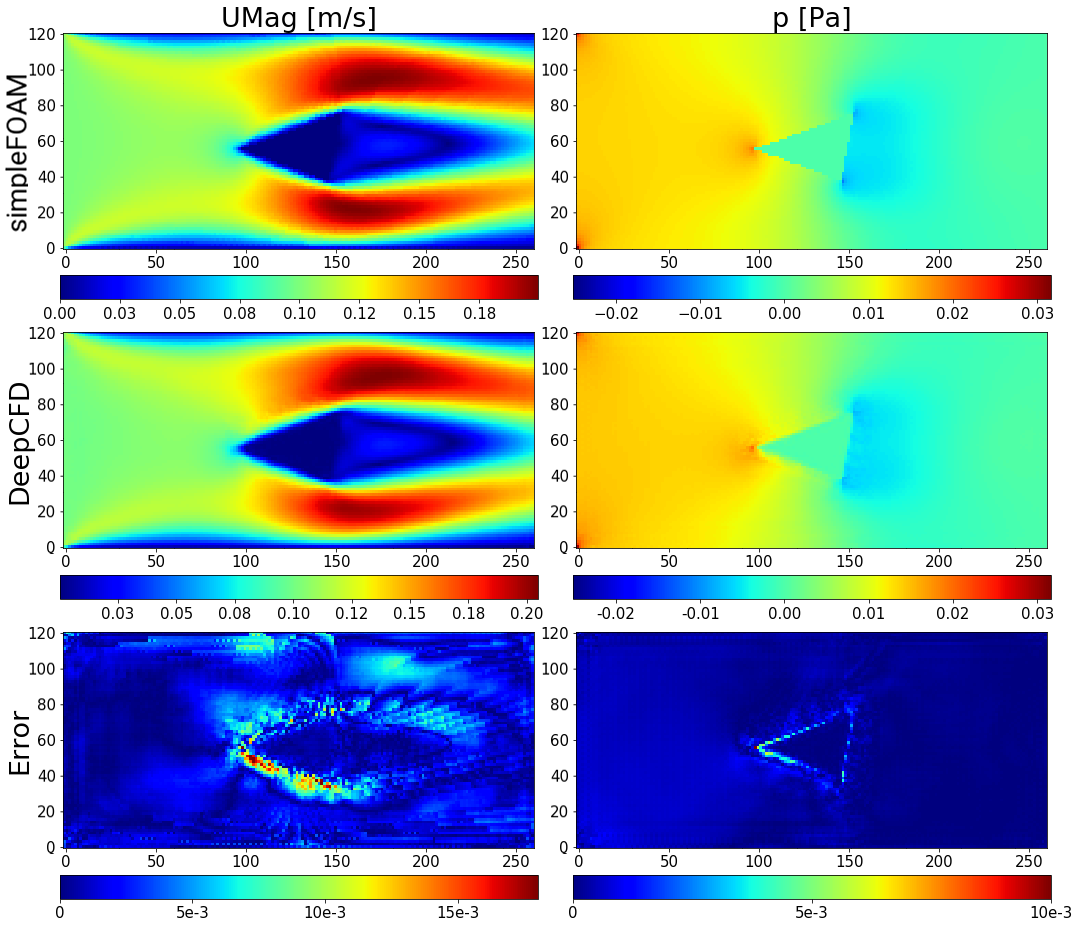}
%\vskip -0.1in
\caption{Comparison between ground-truth CFD (simpleFOAM) and DeepCFD prediction, showing velocity magnitude, pressure, and absolute error in flow around forward-facing triangle shape.}
\label{fig:fig9}
\end{center}
%\vskip -0.2in
\end{figure}

Next plots, in Figures~\ref{fig:fig8}-\ref{fig:fig9}, include the absolute error information between the ground-truth and the machine learned results. Due to space limitation, the velocity magnitude is shown instead of the separate components, and plots positions have been rearranged. The first two rows show, respectively, ground-truth CFD and DeepCFD data, whereas the last row shows the absolute error for both velocity magnitude (first column) and pressure (second column). For complete plots of this and other flows with separate velocity components, the reader can refer to the supplementary material provided with the paper.

Figure~\ref{fig:fig8} shows the flow around a circle based shape, which creates a round region of high pressure on its leading edge, and flow separation happens as the fluid approaches the middle of the obstacle from both top and bottom surfaces. In Figure~\ref{fig:fig9}, the forward-facing triangle shape forms a considerably smaller region of high pressure at its frontal vertex, with flow separation occurring only further downhill. The DeepCFD model is able to correctly capture all these phenomena at a cost of very low error rates. 

\subsection{Quantitative Analysis}
\label{sec44}

\begin{figure}[ht!]
\begin{center}
\centerline{\includegraphics[width=1\columnwidth]{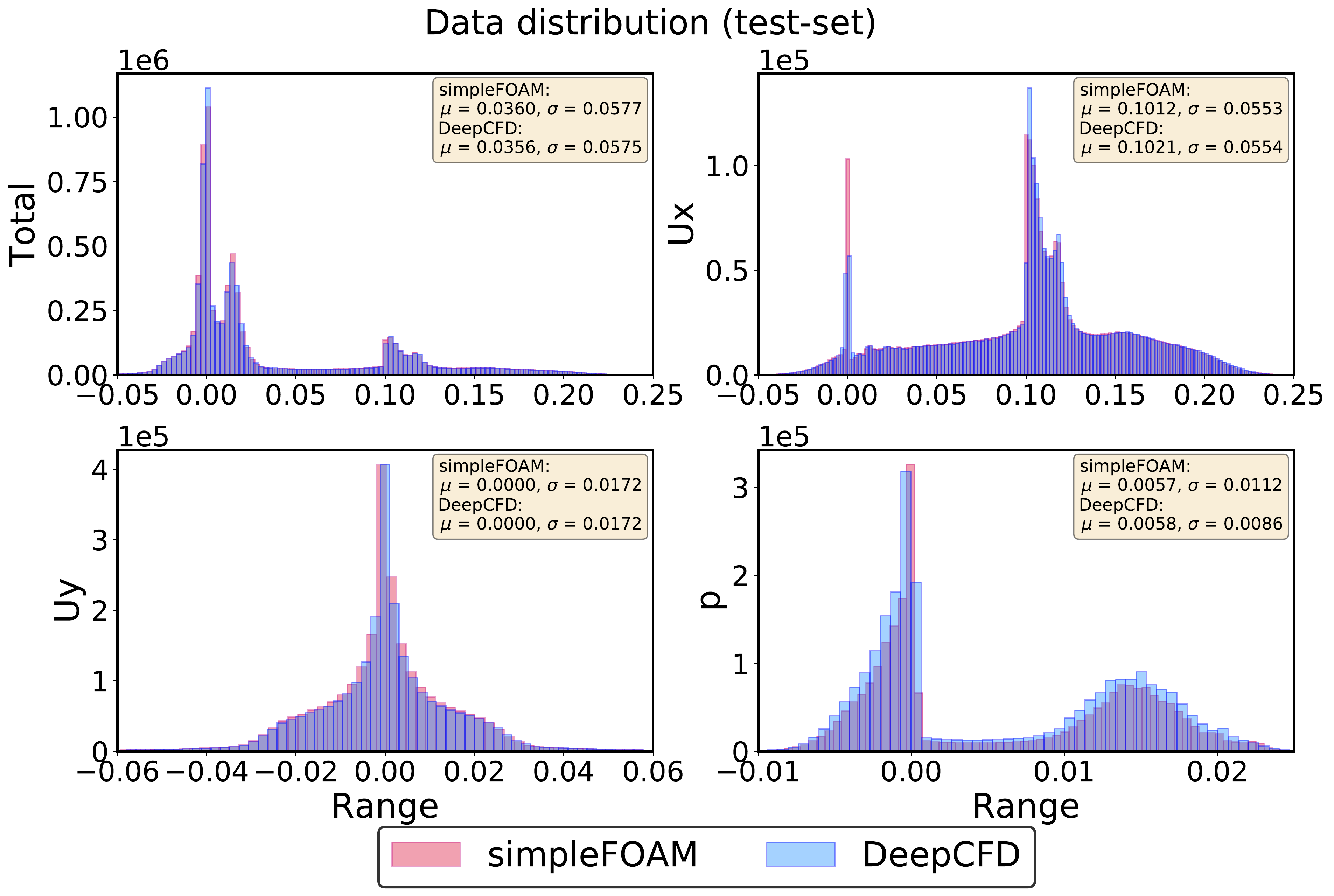}}
\caption{Ground-truth CFD data distribution against DeepCFD predicted distribution from 295 test samples.}
\label{fig:fig10}
\end{center}
%\vskip -0.2in
\end{figure}

In Figure~\ref{fig:fig10}, the ground-truth CFD data distribution is plotted against the DeepCFD modeled data distribution from 295 test samples. For consistency with previously shown figures, the first plot on the top left presents the combined data with all variables (Ux, Uy, and p), whereas the other plots show the data distribution for each specific variable.

\begin{figure}[ht!]
\begin{center}
\centerline{\includegraphics[width=1\columnwidth]{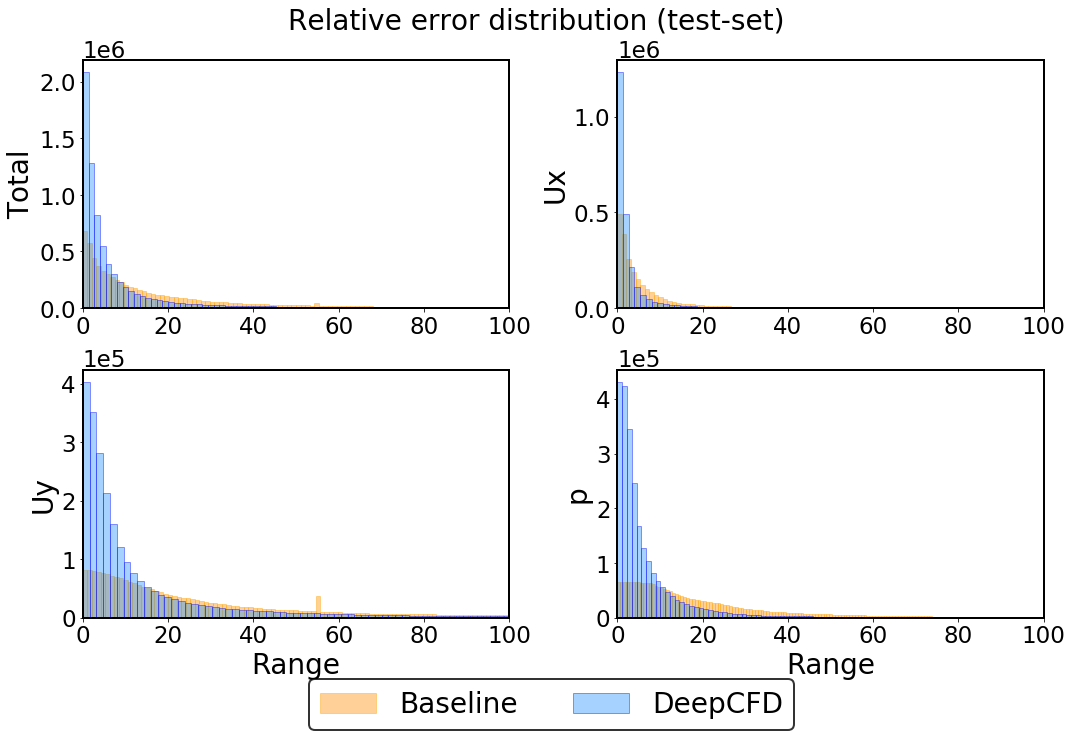}}
\caption{Relative error distribution for predictions using baseline (orange) and DeepCFD (blue) models on 295 test samples.}
\label{fig:fig11}
\end{center}
\vskip -0.2in
\end{figure}

In agreement with the qualitative plots formerly presented, the approximated DeepCFD solution on the test-set produces data distributions with shapes very similar to the ones from the ground-truth CFD simulation for all quantities analysed. Furthermore, mean and standard deviation differences between the two approaches are very small. In order to evaluate how these deviations from DeepCFD compare to those obtained with the baseline, Figure~\ref{fig:fig11} shows the relative error distributions for each of these models.

The relative error is given by $\abs{ \frac{p - gt}{gt+k} } \times 100 \%$, where $p$ is the prediction, and $gt$ is the ground-truth value. Because this metric fails when $gt$ approaches zero (division by zero), an adjusting scalar $k = \num{1e-4}$ is plugged in the denominator to address the issue. As shown, the predictions made by the proposed model tend to be concentrated on the lower end of relative errors (most values with less than 10\% error), whereas the baseline shows a wider error distribution with considerably more elements with higher error rates.

Finally, the performance of DeepCFD, in terms of prediction time, is tested and compared against the standard CFD solution in Table~\ref{tab:table3}. Since the steady-state flow solver used here is not implemented for GPU runs, the reference time of the standard CFD approach was taken from the average of 50 random runs on one single core of the Intel Xeon E-2146G processor. In the case of the machine learned approach, approximations can be generated using both CPUs and GPUs. Therefore, the time taken for DeepCFD to make a prediction was evaluated on the same CPU used to generate the ground-truth CFD data, as well as on an Nvidia Geforce RTX-2080 Ti GPU. Moreover, the time results were averaged from 1000 different runs using three different batch sizes: 1, 10, and 100. Average and standard deviation of run times, as well as speedup values are provided.

\begin{table}[ht!]
\caption{Run time and speedup comparisons.}
\label{tab:table3}
\vskip 0.15in
\begin{center}
\begin{small}
\begin{sc}
\begin{tabular}{lcc}
\toprule
& CFD (CPU) & \\
Batch size  &  Time (s)  &  Speedup \\
\midrule
1   &    \text{52.51 $\pm$ 15.27 } & -    \\
\midrule
 & DeepCFD (CPU) & \\
Batch size  &  Time (s)  &  Speedup \\
\midrule
1    & \text{$\num{4.77e-2}$ $\pm$ $\num{7.15e-4}$ } & $\num{1.10e3}$  \\  
10    & \text{$\num{3.57e-2}$ $\pm$ $\num{5.44e-4}$ } &   $\num{1.47e3}$  \\
100    & \text{$\num{3.50e-2}$ $\pm$ $\num{7.07e-4}$ } &    $\num{1.50e3}$ \\
\midrule
&  DeepCFD (GPU)  & \\
Batch size  &  Time (s)  &  Speedup \\
\midrule
1    & \text{$\num{4.57e-3}$ $\pm$ $\num{1.03e-4}$ } & $\num{1.15e4}$  \\
10    & \text{$\num{6.81e-4}$ $\pm$ $\num{1.03e-4}$ } &   $\num{7.71e4}$  \\
100    & \text{$\num{1.02e-4}$ $\pm$ $\num{1.03e-4}$ } &    $\num{5.14e5}$ \\
\bottomrule
\end{tabular}
\end{sc}
\end{small}
\end{center}
\vskip -0.1in
\end{table}

For a fair evaluation, only the CPU-CPU time comparison can be considered a speedup advantage, since GPU reference time measurements are not available. Moreover, due to the characteristics of the pressure-velocity coupling, CFD solution times may vary considerably, thus the average solution time was considered. Even in this case, up to three orders of magnitude speedup was obtained for all batch sizes used. Increasing the batch-size improved the performance only marginally for the CPU runs. However, because machine learning models can be easily run on GPUs, the GPU evaluation times were also considered. In that scenario, prediction times can vary considerably depending on the batch size, starting with 4 orders of magnitude speedup with batch size 1 and up to 5 orders of magnitude with 100 samples.

\section{Conclusions and Future Work}
\label{sec5}

In this paper, we proposed a new and efficient way for approximating non-uniform steady laminar flow CFD calculations. Previous works that addressed this problem could provide only a solution for the velocity field, but we showed that an U-Net architecture can be employed to provide complete solutions of coupled velocity and pressure fields. The ground-truth CFD data was created with an extensively validated numerical solver and a workflow for generating training samples for a channel flow around randomly shaped objects was developed. After comprehensive hyper-parameter search, considering 4 different architectures in 108 parameter configurations, we found that the proposed DeepCFD model (U-Net architecture using separate decoders) outperformed all other models, including the baseline (Autoencoder with 3 decoders). The proposed DeepCFD model with optimum parameters was then used in additional experiments in order to perform further analyses in both qualitative and quantitative terms. 

Relevant discussion about the model's accuracy and performance in comparison to the baseline and the standard CFD approach was provided. At a cost of low error rates, a speedup of up to 3 orders of magnitude can be achieved (CPU-CPU), or even of up to 5 orders of magnitude (GPU-CPU). Finally, we provide code and dataset as supplementary material in order to contribute with further expansion of the field of data-driven models for CFD. 

For future work, we intend to extend the 2D methodology used here for 3D flow configurations, as well as expand the dataset in terms of number of samples and variability for more complex generalization in real flow conditions. Furthermore, since most flows of interest for engineering are turbulent, we intend to incorporate the model developed here to recurrent neural networks architectures in order to take the time dependency into account. Most efforts in this field concentrate on developing closure terms for governing equations from global turbulent quantities, but the ability of CNNs in learning spatial characteristics of the flow can also be leveraged in such future investigations. Finally, we also intend to incorporate physical constraints to the neural network training procedure, so that the network prediction is bounded within the physical constraints of the problem.

%\section*{Acknowledgments}

%\section*{References}

\bibliography{bib}
\bibliographystyle{elsarticle-num-names}

\end{document}